\documentstyle[aps]{revtex}

\begin{document}
\draft
\title{Driven Dynamics: A Probable Photodriven Frenkel-Kontorova Model}
\author{Bambi Hu$^{1,4}$ and Jian-Yang Zhu$^{1,2,3}\thanks{%
Author to whom correspondence should be addressed. Email address:
zhujy@public.nc.jx.cn}$}
\address{$^1$Centre for Nonlinear Studies and Department of Physics, Baptist
University, Hong Kong\\
$^2$Department of Physics, Jiangxi Normal University, Nanchang, 330027,
China \\
$^3$Department of Physics, Beijing Normal University, Beijing, 100875, China%
\\
$^4$Department of Physics and Texas Center for Superconductivity, University
of Houston, Houston, TX 77204, USA}
\maketitle

\begin{abstract}
In this study, we examine the dynamics of a one-dimensional
Frenkel-Kontorova chain consisting of nanosize clusters (the ''particles'')
and photochromic molecules (the ''bonds''), and being subjected to a
periodic substrate potential. Whether the whole chain should be running or
be locked depends on both the frequency and the wavelength of the light
(keeping the other parameters fixed), as observed through numerical
simulation. In the locked state, the particles are bound at the bottom of
the external potential and vibrate backwards and forwards at a constant
amplitude. In the running state, the initially fed energy is transformed
into directed motion as a whole. It is of interest to note that the driving
energy is introduced to the system by the irradiation of light, and the
driven mechanism is based on the dynamical competition between the inherent
lengths of the moving object (the chain) and the supporting carrier (the
isotropic surface). However, the most important is that the light-induced
conformational changes of the chromophore lead to the time-and-space
dependence of the rest lengths of the bonds.
\end{abstract}


The driven dynamics of a system of interacting particles in atomic or
mesoscopic scale has been attracting much attention. It is very important in
technology, very rich in physics, and widely applicable in many fields, such
as mass transport, conductivity, tribology, etc. As far as we know, one of
the most typical examples conducted on the driven dynamics has been the
driven Frenkel-Kontorova-type systems biased by dc, and ac external forces,
respectively.

The Frenkel-Kontorova (FK) model\cite{FKmodel} may describe, for example, a
closely packed row of atoms in crystals, a layer of atoms adsorbed on
crystal surface, a chain of ions in a ''channel'' of quasi-one-dimensional
conductors, hydrogen atoms in hydrogen-bonded systems, and so on. In
general, such a system can be treated with two parts in the driven dynamical
model: the moving object and the supporting carrier. First, the moving
object is considered an atomic subsystem, in which the interparticle
interaction is taken as a harmonic interaction between the nearest
neighbors. Second, the action of the supporting carrier to the moving object
is modeled as an external potential, a damping constant, and a thermal bath.
When, for instance, an external dc driving force $f$ is applied to such a
system, its response can be very nonlinear and complex. The overdamped case (%
$\eta \gg \omega _0$, where $\eta $ is the external damping and $\omega _0$
is the vibrational frequency at the bottom of the periodic potential), has
been studied in a number of papers \cite{Over1,Over2,Over3}. When the
driving force $f$ changes, the multistep and the hysteretic transition of
the system from the locked state to the sliding state in the underdamped
case ($\eta \ll \omega _0$) has also been delineated in detail\cite
{under1,under2}. At the same time, various intermediate regimes can be
described by resonance phenomena\cite{intermediate1} and by the moving
quasiparticle excitation, kinks\cite{intermediate2}.

A theoretical demonstration of driven dynamical behavior, as a steady-state
response to an applied driving force, must be well verified in laboratory.
It is of interest to note that a Josephson junction array is just an
technically and physically excellent demonstration, of many typical
nonlinear features of the driven FK-type systems biased by, for instance, dc
forces\cite{dc} and ac forces\cite{ac}.

In this paper, we propose a new photo-driven Frenkel-Kontorova model, which
may be actualized in experiment. The basic challenge of our proposal is to
keep the time-and-space dependence of the rest lengths of the harmonic
oscillator, as was suggested originally by Proto {\it et al}.\cite{Porto}
when discussing the atomic scale engines. In this model system, the
supporting carrier is taken as an isotropic surface, and the moving object
(a FK chain) consists of nanosize clusters (the ''particles'') and
photochromic molecules\cite{molecules} (the ''bonds''). The particles are
identical with mass $m$; the bonds are flexible with elastic strength $k$
and photoactive with sensitive coefficient $b$; the time-and-space
dependence of the free equilibrium rest lengths is due to the light-induced
conformational changes of the chromophore. In contrast to the particles, the
bond's mass is so small that it can be neglected. For simplicity we restrict
the discussion to one dimension at zero temperature. The $N$ equations of
motion read as 
\begin{equation}
m\ddot{x}_n+\eta \dot{x}_n+\frac{\partial \Phi \left( x_n\right) }{\partial
x_n}+\sum_\delta \frac{\partial \Psi \left( x_n,x_{n+\delta },a_{n,n+\delta
}\left( t\right) \right) }{\partial x_n}=0,  \label{motion}
\end{equation}
where $x_n$\ is the coordinate of the $n$th particle with $1\leq n\leq N$.
The second term of Eq.(\ref{motion}) describes the dissipative interaction
(damping) between the particles and the surface. It is proportional to the
relative velocities of the particles, with proportionality constant $\eta $.
The static interaction between the particles and the surface is represented
by the periodic substrate potential 
\begin{equation}
\Phi \left( x_n\right) =\Phi _0\left( 1-\cos \frac{2\pi x_n}{\lambda _1}%
\right) .  \label{external potential}
\end{equation}
Concerning the interparticle interaction, we take the form of the nearest
neighboring harmonic interaction 
\begin{equation}
\Psi \left( x_n,x_{n+\delta },a_{n,n+\delta }\left( t\right) \right) =\frac k%
2\left[ \left| x_n-x_{n+\delta }\right| -a_{n,n+\delta }\left( t\right)
\right] ^2.  \label{harmonic interaction}
\end{equation}
Due to the light sensitivity, the bond lengths change with the irradiation
of light. The free equilibrium rest lengths $a_{n,n+\delta }\left( t\right) $
depend on both the bond's position, specified by the indices $n$, $n+\delta $%
, ($\delta =\pm 1$), and time $t$. We restrict ourselves to a certain choice
for $a_{n,n+\delta }\left( t\right) $: 
\begin{eqnarray}
a_{n,n+\delta }\left( t\right) &=&a_0\left( 1+b\cos \left( qx_{n,n+\delta
}-\omega t\right) \right)  \nonumber \\
&=&a_0\left[ 1+b\cos \left( \frac{2\pi }{\lambda _2}\frac{x_n+x_{n+\delta }}2%
-\omega t\right) \right] ,  \label{length}
\end{eqnarray}
where $b$ $\left( b<1\right) $ is an absorption coefficient (amplitude) less
than one, $a_0$\ is the rest length of the harmonic interaction, and $%
\lambda _2$ and $\omega $ are the wavelength and frequency of external
irradiated light, respectively. Obviously, $a_0$ and $b$ are intrinsic
parameters, and $\lambda _2$ and $\omega $ are externally adjustable. Light
energy, if pumped into the system in an irradiating manner, may produce
spatially and temporally correlated changes of the lengths $a_{n,n+\delta
}\left( t\right) $. Then, the dynamical local competition between the
periodicity $\lambda _1$ and the rest lengths $a_{n,n+\delta }\left(
t\right) $ may induce a driving power, which will impel the chain into
motion in some manner. Granted, this power must be strong enough to overcome
the resistance.

After introducing dimensionless parameters $x_n\rightarrow \left( 2\pi
/\lambda _1\right) x_n$, $t\rightarrow \left( 2\pi /\lambda _1\right) \sqrt{%
\Phi _0/m}t$, and the renormalized parameters $\omega \rightarrow \left[
\left( 2\pi /\lambda _1\right) \sqrt{\Phi _0/m}\right] ^{-1}\omega $, $\eta
\rightarrow \left[ \left( 2\pi /\lambda _1\right) \sqrt{m\Phi _0}\right]
^{-1}\eta $, $k\rightarrow \left( 1/\Phi _0\right) \left( \lambda _1/2\pi
\right) ^2k$, $a_0\rightarrow \left( 2\pi /\lambda _1\right) a_0,$ the
motion equation reads as follows: 
\begin{eqnarray}
\ddot{x}_n &=&-\eta \dot{x}_n-\sin x_n  \nonumber \\
&&+k\left[ x_{n+1}-2x_n+x_{n-1}+2a_0b\sin \left( \frac{\lambda _1}{\lambda _2%
}\frac{x_{n+1}-x_{n-1}}4\right) \sin \left( \frac{\lambda _1}{\lambda _2}%
\frac{x_{n+1}+2x_n+x_{n-1}}4-\omega t\right) \right]  \nonumber \\
&&-\frac{ba_0}2\frac{\lambda _1}{\lambda _2}k\left[ \left(
x_n-x_{n-1}-a_0\right) \sin \left( \frac{\lambda _1}{\lambda _2}\frac{%
x_n+x_{n-1}}2-\omega t\right) -\frac{ba_0}2\sin \left[ 2\left( \frac{\lambda
_1}{\lambda _2}\frac{x_n+x_{n-1}}2-\omega t\right) \right] \right]  \nonumber
\\
&&-\frac{ba_0}2\frac{\lambda _1}{\lambda _2}k\left[ \left(
x_{n+1}-x_n-a_0\right) \sin \left( \frac{\lambda _1}{\lambda _2}\frac{%
x_{n+1}+x_n}2-\omega t\right) -\frac{ba_0}2\sin \left[ 2\left( \frac{\lambda
_1}{\lambda _2}\frac{x_{n+1}+x_n}2-\omega t\right) \right] \right] . 
\nonumber \\
&&  \label{dimensionless}
\end{eqnarray}
In the present survey, the mass of the particle is $m=1$, the period and
amplitude of the external sinusoidal potential are $\lambda _1=2\pi $ and $%
\Phi _0=1$, respectively. So, if we take the natural unit like this, i.e.,
the time scaled by $\left( 2\pi /\lambda _1\right) \sqrt{\Phi _0/m}$, the
length scaled by $\left( 2\pi /\lambda _1\right) $, and the mass scaled by $%
\left( 1/m\right) $, then the physical quantities, time, velocity and so on,
become dimensionless. And one should, in order to restore them, multiply
spatial variables by $\left( \lambda _1/2\pi \right) $, times by $\left(
\lambda _1/2\pi \right) \sqrt{m/\Phi _0}$, mass by $m$, and energies by $%
\Phi _0$, etc.

In our model, parameters $\eta $, $k$, $b$, $\lambda _2$, $\omega $ are all
very important. For example, when the external viscous damping coefficient $%
\eta $ is overdamped $\eta \gg \omega _0$, underdamped $\eta \ll \omega _0$,
or other intermediate regions, the{\it \ }driven system exhibits very
different dynamical behavior, which has been studied in a number of papers
as mentioned above. Note that in the present survey, the characteristic
frequency of particle vibration at the minimum of the external potential is $%
\omega _0=1$. To simplify the problem, we keep the damping coefficient and
the elastic strength fixed as $\eta =0.1$ and $k=1$. Here, the value of $k$
also corresponds to the intermediate case between a strong coupled
(Sine-Gordon) and a weakly coupled chain. Thus, we focus on the parameters $%
b,$ $\lambda _2$ and $\omega $, all of which definitely characterize how the
free equilibrium rest lengths $a_{n,n+\delta }\left( t\right) $ depend on
the time and position of the bonds.

Our simulation of the Runge-Kutta iteration method starts when the particles
are at the bottom of the substrate potential well at rest, and the boundary
condition is chosen to be the simplest, as $x_{j+N}=x_j+2\pi N$ (periodic
boundary condition). In this way, $a_0=\lambda _1=2\pi $. Only a finite
system possible, its size should be appropriate to ensure that $M$ waves ($M$
is the wave-number) can be inserted into the background structure of $N$%
-particles system; the integers $N$ and $M$ must satisfy the relation: $%
N/M=\lambda _2/\lambda _1$. In all calculations, we took the natural unit ($%
m=1$, $\lambda _1=2\pi $ and $\Phi _0=1$) and kept $\eta =0.1$ and $k=1,$ as
mentioned above.

In Fig.1 we show the velocity of the center of mass of the system $v_c$ [$%
v_c=\left( \sum_{i=1}^Nm_i\dot{x}_i\right) /\left( \sum_{i=1}^Nm_i\right)
=\left( \sum_i\dot{x}_i\right) /N$, $m_i=m$] as a function of time $t$ for
some different parameters. The results tell us the following: (1) After
passing through a transitory relaxation, the system turns quickly into a
steady-state; (2) There are two types of steady-states, the locked state
(Fig.1(a)-(c)) and the running state (Fig.1(d)-(e)). In the locked state,
the particles are bound at the bottom of the substrate potential and vibrate
backwards and forwards at a constant amplitude. In the running state, the
initially fed energy is transformed into directed motion as a whole, with
almost constant speed except for a little undulation. With different
parameters, there are different oscillating amplitude (in the locked state)
or running speed (in the running state). On the whole, the system exhibits
interesting and parameter-dependent dynamical behavior.

The amplitude $b$ intrinsically characterizes the light sensitivity of the
photochromic molecules (bonds) and determines the elastic energy stored. To
find more about it, we started from $b=0$, slowly increased it by small
steps $\Delta b=0.01$. With every $b$ value, we let the system relax to $%
10^4 $ time steps, $10^{-3}$ each, to ensure a stationary state. We then
calculated the average velocity of the centre of mass $\left\langle
v_c\right\rangle $ over the time period of $10^4\sim 10^6$ steps. The
results for $\lambda _2/\lambda _1=2.5$, $3$ and $\omega =1$, $3$,
respectively, are shown in Fig.2 (a) and (b). The processes both have a $%
\left\langle v_c\right\rangle $ increasing from zero as soon as the
amplitude exceeds a critical value $b_c$. At zero speed, each particle is
locked in one external potential well without any observable macroscopic
motion, expecting its emancipation as the driving energy exceeds a certain
value. This is easy to understand for a dissipation system.

It is no surprise as well, that each particular situation is unique, given
different frequencies and wavelengths as temporal and spatial modulations.
To get further information of their effects, our next simulation was focused
on the external photo-driving parameters. Figure 3 presents $\left\langle
v_c\right\rangle $ as a function of the driving frequencies $\omega $ for a
fixed value of the wavelength $\lambda _2$. From these numerical results,
three things are note-worthy. First, as the magnitude of frequency slowly
increases from $\omega =0$, a dynamical transition occurs, from an initially
locked state to a running state, and finally back to a locked state again.
Second, there exits a peak of maximal velocity. Third, as the wavelength
increases, the running state has a lower maximal velocity and exists in a
narrower span of frequency with lower values. The same is true for $%
\left\langle v_c\right\rangle $ as a function of wavelength, as is presented
in Fig. 4.

Finally, we present a dynamical phase diagram in Fig.5, where we plot the
driving frequency versus the ratio of the optical wavelength to the period
of the substrate potential. The figure shows that, within a finite range of
driving frequencies and wavelengths, the system as a whole will be running;
otherwise, the particles will be locked at the bottom of the substrate
potential, permanently.

In conclusion, we would like to emphasize several points. First, a
reasonable explanation for the photo-driven FK model seems to be that
temporal and spatial modulations both result in the resonant absorption of
optical energy. And, the dynamical competition between the inherent lengths
of the moving object (the chain) and the supporting carrier (the isotropic
surface) leads to the transformation of fed energy to directed motion.
Second, the light-induced conformational changes of the photochromophore
leads to the time-and-space dependence of the rest lengths of the bonds,
which is the most important in this model. Third, according to our
calculations, the steady state of the system is not affected by the initial
conditions. The drift direction of the chain lies only on the sign in front
of $\omega $ in the expression $a_{n,n+\delta }\left( t\right) =a_0\left(
1+b\cos \left( qx_{n,n+\delta }\mp \omega t\right) \right) $. Namely, ''$-$%
'' and ''$+$'' correspond to a motion in positive and negative $x$%
-direction, respectively. Therefore, the motion of the system can be easily
controlled. Fourth, the ''springs'' connecting the ''particles'' are
light-sensitive molecules (photochromic molecules) and can expand or shrink
with light reversibly\cite{photochromism}; adjusting the frequency or the
wavelength of the lightwave may produce spectacular motions. We believe that
the concept presented in this paper is simple and robust enough to be
realized in actual experiments, and may demonstrate many interesting and
typical nonlinear features of a system of interacting particles. However, we
would like to further state that the connection between the model and
photochromic chains stands at the level of speculation, being an interesting
open question or idea, still to be proved.

\acknowledgments{Discussions with all members of the Center for Nonlinear
Studies at HKBU are gratefully acknowledged. }This work was supported in
part by grants from the Hong Kong Research Grants Council (RGC) and the Hong
Kong Baptist University Faculty Research Grant (FRG). J.Y.Z was also
supported from the National Basic Research Project ``Nonlinear Science'',
and the National Natural Science Foundation of China under Grant No.
10075025.

\null\vskip 0.2cm

\centerline{\bf Caption of figures} \vskip1cm

Fig.1: Velocity of the centre of mass $v_c$ of chain as a function of time $%
t $. For different parameters, the system turns quickly into a steady-going
locked state ((a)-(c)) or a steady-going running state ((d),(e)) passing
through a transitory relaxation.

Fig.2: Average velocity of the centre of mass $<v_c>$ of chain as a function
of the amplitude $b$.

Fig.3: Average velocity of the centre of mass $<v_c>$ of chain as a function
of the driving frequency $\omega $.

Fig.4: Average velocity of the centre of mass $<v_c>$ of chain as a function
of $\lambda _2/\lambda _1$ that denotes the ratio of the optical wavelength
to the period of the substrate potential. (Although the relation $%
N/M=\lambda _2/\lambda _1$ is not satisfied everywhere, the result is still
credible with the large system size of $N=1000$.)

Fig.5: Dynamical phase diagram in the ($\omega ,\lambda _2/\lambda _1$)
plane for the photo-driven one-dimensional FK model. Here parameters are
fixed at $b=0.3,$ $\eta =0.1$ and $k=1,$ respectively. And the
particle-number $N$ is chosen in $450\sim 550$ to ensure that $M$ waves ($M$
is an integer, $N$ and $M$ satisfy the relation: $N/M=\lambda _2/\lambda _1$%
) can be inserted into the commensurate background structure of $N$-particle
system from first to last.


\begin{references}
\bibitem{molecules}  Light-induced reversible isomerizations between two
forms having different absorption spectra are referred to photochromism, and
the compounds capable of these reactions are called photochromic molecules.
Diarylethenes are known as the new class of thermally irreversible and
fatigue resistant photochromic molecules.
\end{references}
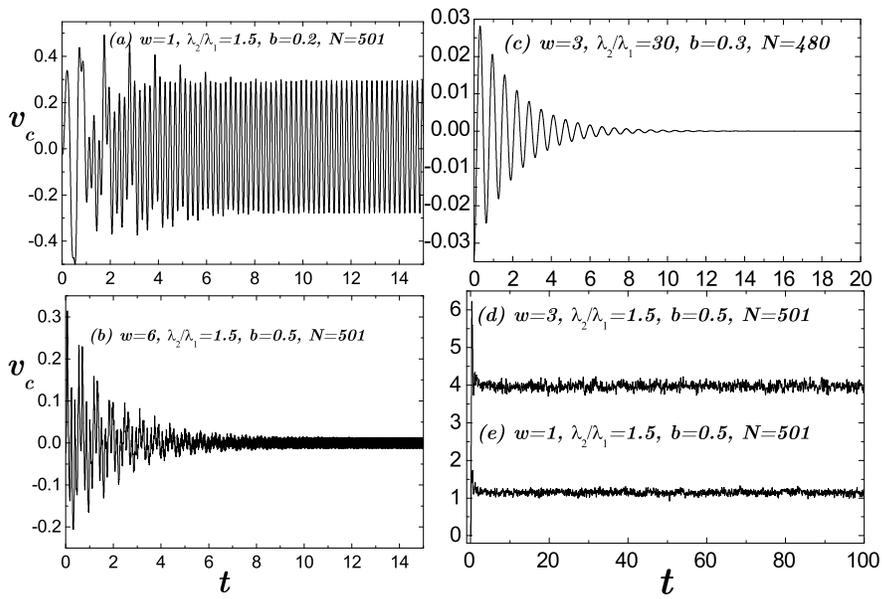
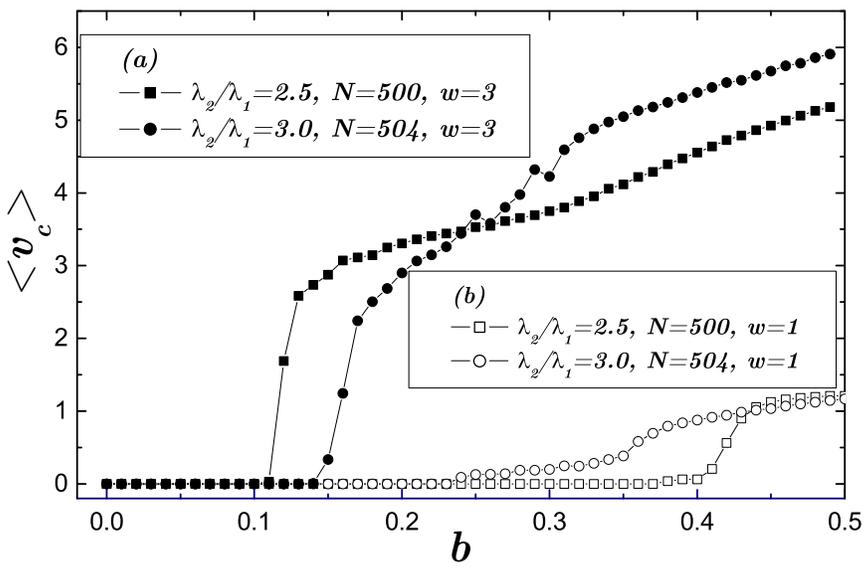
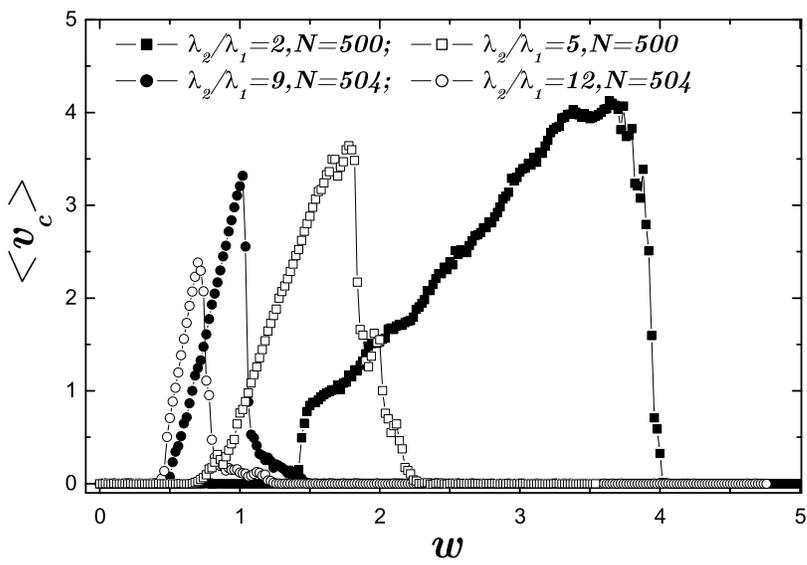
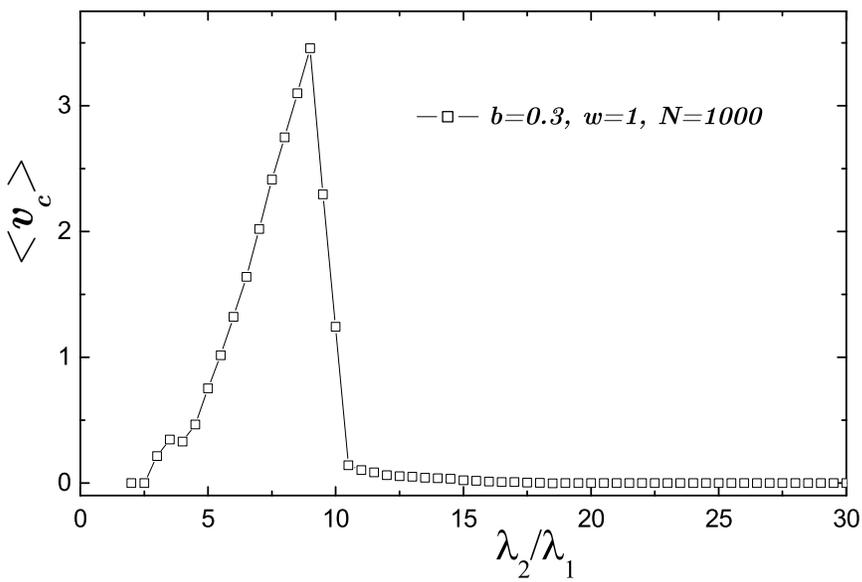
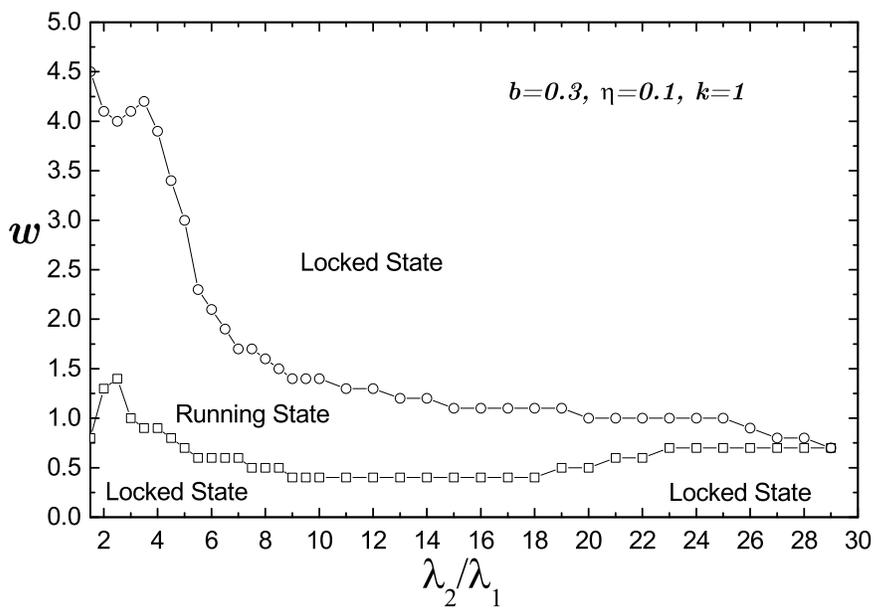
\end{document}